\documentclass[12pt]{article}
\usepackage{graphicx}
\usepackage{amsfonts}
\usepackage{amssymb,amsmath}
\usepackage{color}
\usepackage[colorlinks=true,linkcolor=blue,citecolor=blue]{hyperref}
\usepackage{epsf}

\begin{document}

\begin{titlepage}

\begin{flushright}
IPMU 12-0195\\
ICRR-Report-634-2012-23
\end{flushright}

\begin{center}

{\Large \bf  
CMB constraint on non-Gaussianity in isocurvature perturbations}

\vskip .45in

{
Chiaki Hikage$^{a}$,
Masahiro Kawasaki$^{b,c}$,
Toyokazu Sekiguchi$^d$ and
Tomo Takahashi$^e$
}

\vskip .45in

{\em
$^a$Kobayashi-Maskawa Institute, Nagoya University, Nagoya 464-8602, Japan \vspace{0.2cm}\\
$^b$Institute for Cosmic Ray Research,
University of Tokyo, Kashiwa 277-8582, Japan \vspace{0.2cm}\\
$^c$ Kavli Institute for the Physics and Mathematics of the Universe,
University of Tokyo, Kashiwa 277-8568, Japan \vspace{0.2cm}\\
$^d$Department of Physics and Astrophysics, Nagoya University, Nagoya 464-8602, Japan\\
$^e$Department of Physics, Saga University, Saga 840-8502, Japan
}

\end{center}

\vskip .4in

\begin{abstract}

We study the CMB constraints on non-Gaussianity in CDM isocurvature perturbations.
Non-Gaussian isocurvature perturbations can be produced in various models 
at the very early stage of the Universe. Since the isocurvature perturbations little affect the
structure formation at late times, CMB is the best probe of isocurvature non-Gaussianity 
at least in the near future. In this paper, we focus on 
non-Gaussian curvature 
and isocurvature perturbations of the local-type, which are uncorrelated and in the form
$\zeta=\zeta_{\rm G}+\frac35f_{\rm NL}(\zeta_{\rm G}^2-\langle \zeta_{\rm G}^2\rangle)$
and $S=S_{\rm G}+f^{({\rm ISO})}_{\rm NL}(S_{\rm G}-\langle S_{\rm G}^2\rangle)$, 
and constrain the non-linearity parameter of isocurvature perturbations, $f^{({\rm ISO})}_{\rm NL}$, 
as well as the curvature one $f_{\rm NL}$.
For this purpose, we employ several state-of-art 
techniques for the analysis of CMB data and simulation. 
Assuming that isocurvature perturbations are subdominant, 
we apply our method to the WMAP 7-year data of temperature anisotropy and 
obtain constraints on a combination $\alpha^2 f^{({\rm ISO})}_{\rm NL}$,
where $\alpha$ is the ratio of the power spectrum of isocurvature perturbations
to that of the adiabatic ones.
When the adiabatic perturbations are assumed to be Gaussian, 
we obtained a constraint $\alpha^2 f_{\rm NL}^{\rm (ISO)}=40\pm66$
assuming the power spectrum of isocurvature perturbations is scale-invariant.
When we assume that the adiabatic perturbations can also be non-Gaussian, 
we obtain $f_{\rm NL}=38\pm24$ and $\alpha^2 f_{\rm NL}^{\rm (ISO)}=-8\pm72$. 
We also discuss implications of our results 
for the axion CDM isocurvature model.

\end{abstract}

\end{titlepage}

\setcounter{page}{1}

\section{Introduction} 
\label{sec:introduction}

The adiabaticity, or isocurvature mode of primordial density fluctuations is one of important probes 
of cosmology in various respects. 
Although current cosmological observations suggest that primordial density fluctuations are almost adiabatic and they give severe constraints on 
the size of isocurvature fluctuations,  some fraction of their contribution is still allowed \cite{Komatsu:2010fb}.
Residual isocurvature fluctuations can be generated when there exist multiple components with different origins, 
which are associated with dark matter, baryon and neutrino \cite{Bucher:1999re}\footnote{
Isocurvature fluctuations  in dark radiation have also been studied in Refs. \cite{Kawasaki:2011rc,Kawakami:2012ke}.
}. 
Such examples include axion \cite{Axenides:1983hj,Seckel:1985tj,Linde:1985yf,
Linde:1990yj,Turner:1990uz,Linde:1991km,Lyth:1991ub,Kasuya:2009up} and
Affleck-Dine baryogenesis \cite{Enqvist:1998pf,Enqvist:1999hv,
Kawasaki:2001in}, where
cold dark matter (CDM) and baryon isocurvature modes can be respectively generated.
These isocurvature modes are basically uncorrelated with adiabatic ones, however, 
when one considers a scenario where a light scalar field other than the inflaton is responsible 
for (adiabatic) density fluctuation such as the curvaton model \cite{Enqvist:2001zp,Lyth:2001nq,Moroi:2001ct}, isocurvature perturbations 
can be correlated with the adiabatic ones and be easily generated, depending on how and when CDM and baryon 
are created  \cite{Moroi:2002rd,Lyth:2002my,Moroi:2002vx,Lyth:2003ip,Hamaguchi:2003dc,Gordon:2003hw,Ikegami:2004ve,Ferrer:2004nv,Beltran:2008ei,Moroi:2008nn,Lemoine:2009yu,Lemoine:2009is,Takahashi:2009cx}.
In addition to above mentioned examples, 
a variety of models with isocurvature fluctuations has been extensively studied in 
the literature, hence
the information on isocurvature fluctuations would give a lot of insight on 
various aspects of cosmology, particularly on models of dark matter and baryogenesis  as well as 
those of the early Universe.

Although the theoretical works on non-Gaussianity from isocurvature fluctuations  have been relatively well studied \cite{Kawasaki:2008sn,Langlois:2008vk,Kawasaki:2008pa,Kawakami:2009iu,Langlois:2011zz,Langlois:2010fe}, 
observational constraints on them have not been explored much\footnote{
At linear order, there are  a lot of works which study the observational constraints \cite{Moodley:2004nz,Beltran:2004uv,Bean:2006qz,Trotta:2006ww,Keskitalo:2006qv,Kawasaki:2007mb,Sollom:2009vd,Valiviita:2009bp,Li:2010yb,Valiviita:2012ub}.
}. 
In particular, we cannot find any
work investigating this issue by using the actual data except Ref. \cite{Hikage:2008sk}, where the constraint has been studied with Minkowski functionals using the WMAP 3 data,  although there are a few papers
which elaborate its future CMB constraints  \cite{Hikage:2009rt,Langlois:2011hn,Langlois:2012tm}. 
In the light that we now have precise cosmological data to explore non-Gaussianity as seen 
from the counterpart for adiabatic ones, non-Gaussianity in isocurvature fluctuations
 should be pursued more. 
 
In this paper, we investigate the 
optimal constraints on the local-type 
non-Gaussianity in CDM (baryon) isocurvature 
fluctuations from CMB bispectrum estimator. Although, as mentioned above, isocurvature fluctuations 
can be correlated with adiabatic ones in some cases and there are also other kinds of modes such as 
neutrino one, we in this paper present the methodology of our analysis and 
concentrate to report the constraint on the CDM (baryon) uncorrelated type. 
Constraints on other types such as correlated ones and neutrino modes will be 
reported in a forthcoming paper \cite{Hikage:2012tf}.

The organization of this paper is as follows. In the next section, we give 
the formalism to discuss non-Gaussianity in models with isocurvature fluctuations
and also set our notation. In Section~\ref{sec:analysis},  we describe our analysis method to 
obtain the constraint on non-Gaussianity from isocurvature fluctuations. 
Then in Section~\ref{sec:result},  we present our results on the constraint. 
As an application of our results, we consider constraints on 
the axion isocurvature model in Section \ref{sec:axion}.
The final section is devoted to the conclusion  of this paper.

\section{Model of non-Gaussian perturbations and CMB signature} 
\label{sec:NG}

We consider primordial curvature and CDM isocurvature  perturbations, respectively denoted as 
$\zeta$ and $S$,  in the following form: 
\begin{eqnarray}
\zeta(\vec x)&=&\zeta_{\rm G}(\vec x)
+\frac35f_{\rm NL}(\zeta_{\rm G}(\vec x)^2-\langle \zeta_{\rm G}(\vec x)^2\rangle), 
\label{eq:zeta} \\
S(\vec x)&=&S_{\rm G}(\vec x)
+f_{\rm NL}^{\rm (ISO)}(S_{\rm G}(\vec x)^2-\langle S_{\rm G}(\vec x)^2\rangle), 
\label{eq:S}
\end{eqnarray}
where $\zeta_{\rm G}$ and $S_{\rm G}$ are Gaussian parts of the primordial curvature and 
isocurvature perturbations, respectively. $f_{\rm NL}$ and $f_{\rm NL}^{\rm (ISO)}$ are 
the non-linearity parameters of the curvature and isocurvature perturbations, respectively.
In the following, we denote these primordial perturbations with $X^A(\vec x)$.
Then Eqs. \eqref{eq:zeta} and \eqref{eq:S} can be recast into
\begin{equation}
X^A(\vec x)=X^A_{\rm G}(\vec x)
+f_{\rm NL}^A(X^A_{\rm G}(\vec x)^2-\langle X^A_{\rm G}(\vec x)^2\rangle), 
\end{equation}
where $X_{\rm G}^A$ is the Gaussian part of $X^A$.
Here we defined a  non-linearity parameter $f_{\rm NL}^A$,
which is related to the adiabatic and isocurvature ones via $f_{\rm NL}^\zeta=\frac35f_{\rm NL}$ and 
$f_{\rm NL}^S=f_{\rm NL}^{\rm (ISO)}$.

We note that the non-Gaussian primordial perturbation of 
Eq. \eqref{eq:zeta} is of the so-called local-type, which is discussed in
Refs. \cite{Salopek:1990jq,Gangui:1993tt} as well as many other studies.
Eq. \eqref{eq:S} would be a natural extension of this to isocurvature perturbations
and hence the non-Gaussianity we consider in this paper should be regarded 
as an extension of the local-type one to non-adiabatic perturbations.

In this paper, we consider uncorrelated isocurvature perturbations, 
so that $\langle \zeta_{\rm G} S_{\rm G}\rangle=0$. Thus only 
correlation functions which contain either $\zeta$ or $S$ are non-zero.
In terms of the Fourier components of $X^A$, 
the bispectrum from either the primordial curvature or isocurvature 
perturbations is
\begin{equation}
\langle X^A(\vec k_1)X^A(\vec k_2)X^A(\vec k_3)\rangle
=2f_{\rm NL}^A[P_{X^A_{\rm G}}(k_1)P_{X^A_{\rm G}}(k_2)+
(\mbox{2 perms})] (2\pi)^3\delta^{(3)}(\vec k_1+\vec k_2+\vec k_3),
\label{eq:primb}
\end{equation}
where $P_{X^A_{\rm G}}(k)$ is the power spectrum of the Gaussian perturbations $X^A_{\rm G}$.
Here we neglected loop contributions although we take those into account in Section ~\ref{sec:axion}.

Neglecting the secondary non-Gaussianities
arising from the second or higher order cosmological perturbation theory, 
the harmonic coefficients of 
the CMB temperature anisotropy from primordial perturbations $X^A$
can be given as
\begin{equation}
a^A_{lm}=4\pi(-i)^l\int \frac{dk^3}{(2\pi)^3}g^A_l(k)X^A(\vec k)Y^*_{lm}(\hat k), 
\label{eq:alm}
\end{equation}
where $g^A_l (k)$ is the temperature transfer function
for the primordial perturbations $X^A$.
The total CMB anisotropy is the sum of those from the curvature and
isocurvature perturbations, 
i.e. $a_{lm}=a^\zeta_{lm}+a^S_{lm}$.

Since $\zeta$ and $S$ are uncorrelated we need to consider
their polyspectra which contain only either $\zeta$ or $S$.
Then the total polyspectra are the sum of those from each 
perturbation.
Angular power spectrum $C^A_l$, which is defined by 
$\langle a^A_{lm}{a^A_{l'm'}}^*\rangle=C^A_l\delta_{ll'}\delta_{mm'}$, can be given as
\begin{equation}
C^A_l=\frac2\pi
\int dk\,k^2{g^A_l(k)}^2P_{X^A_{\rm G}}(k).
\end{equation}
The reduced bispectrum $b_{l_1l_2l_3}$ is defined by
\begin{equation}
\langle a_{l_1m_1}a_{l_2m_2}a_{l_3m_3}\rangle
=b_{l_1l_2l_3}\mathcal G^{l_1l_2l_3}_{m_1m_2m_3}, 
\end{equation}
where $\mathcal G^{l_1l_2l_3}_{m_1m_2m_3}$ is the Gaunt integral, 
which can be written in terms of the Wigner-3j symbol as 
\begin{equation}
\mathcal G^{l_1l_2l_3}_{m_1m_2m_3}=
\sqrt{\frac{(2l_1+1)(2l_2+1)(2l_3+1)}{4\pi}}
\begin{pmatrix}
l_1&l_2&l_3 \\
m_1&m_2&m_3 
\end{pmatrix}
\begin{pmatrix}
l_1&l_2&l_3 \\
0&0&0
\end{pmatrix}.
\end{equation}
Given a bispectrum of primordial perturbations in Eq. \eqref{eq:primb}, 
the reduced CMB bispectrum from each of the primordial perturbations 
can be given as 
\begin{equation}
b^A_{l_1l_2l_3}=2f_{\rm NL}^A\int dr r^2\left(
\alpha^A_{l_1}(r)\beta^A_{l_2}(r)\beta^A_{l_3}(r)
+(\mbox{2 perms}) 
\right), \label{eq:cmbb}
\end{equation}
where $\alpha^A_l(r)$ and $\beta^A_l(r)$ are
\begin{eqnarray}
\alpha^A_l(r)&=&\frac2\pi\int dk\,k^2g^A_l(k)j_l(kr), 
\label{eq:alpha} \\
\beta^A_l(r)&=&\frac2\pi\int dk\,k^2g^A_l(k)j_l(kr)P_{X^A_{\rm G}}(k).
\label{eq:beta}
\end{eqnarray}
For later convenience, we introduce a normalized bispectrum
\begin{equation}
\hat b^A_{l_1l_2l_3}\equiv b^A_{l_1l_2l_3}/f^A_{\rm NL}.
\end{equation}

The amplitude of isocurvature power spectrum is constrained from 
CMB angular power spectrum.
In Ref. \cite{Komatsu:2010fb}, 
it is shown that the WMAP observation of CMB gives
an upper bound on the fraction of isocurvature power spectrum in
the total one 
\begin{equation}
\alpha \equiv 
\frac{P_S}{P_\zeta}< 0.15
\end{equation}
at 95 \% confidence level.

Note that the bispectrum is proportional to $\alpha^2f_{\rm NL}^{\rm (ISO)}$.
This can be seen from Eqs.~\eqref{eq:cmbb} and \eqref{eq:beta}.
Hence we report our constraint on non-Gaussianity in isocurvature fluctuations for
the combination of $\alpha^2f_{\rm NL}^{\rm (ISO)}$.

\section{Analysis method} 
\label{sec:analysis}

Here we describe our analysis method to derive the constraints on
the non-linearity parameters from CMB data.
Our method is basically the same as the one given in Ref. \cite{Smith:2009jr}.

\subsection{Estimator of non-linearity parameters}
\label{sec:estim}

In the limit of small non-linearity parameter $f^A_{\rm NL}$, the 
effect of the deviation from Gaussianity manifests in the CMB bispectrum, 
so that the optimal estimator can be constructed from the three point function 
of CMB anisotropy \cite{Komatsu:2003iq}.
We adopt the following cubic estimator of non-linearity parameters 
\begin{equation}
\begin{pmatrix}
\hat f^\zeta_{\rm NL} \\
\hat f^S_{\rm NL}
\end{pmatrix}=
\begin{pmatrix}
\langle S_{\rm prim}^\zeta\rangle_{f^\zeta_{\rm NL}=1} &
\langle S_{\rm prim}^\zeta\rangle_{f^S_{\rm NL}=1} \\
\langle S_{\rm prim}^S\rangle_{f^\zeta_{\rm NL}=1} & 
\langle S_{\rm prim}^S\rangle_{f^S_{\rm NL}=1}
\end{pmatrix}^{-1}
\begin{pmatrix}
S_{\rm prim}^\zeta \\
S_{\rm prim}^S
\end{pmatrix},
\label{eq:estim2}
\end{equation}
where an angle bracket with  subscript $f^A_{\rm NL}=1$ indicates 
an ensemble average over simulations with
non-zero non-linearity parameter shown in the subscript with
the other non-linearity parameter being fixed to zero.
$S^A_{\rm prim}$ is the cubic statistics
\cite{Komatsu:2003iq,Yadav:2007rk,
Creminelli:2005hu,Yadav:2007ny}, which is given by
\begin{equation}
S^A_{\rm prim}=\frac16\sum_{\{lm\}}
\hat b^A_{l_1l_2l_3}
\mathcal G^{l_1l_2l_3}_{m_1m_2m_3}
\left[\tilde a_{l_1 m_1}
\tilde a_{l_2 m_2}
\tilde a_{l_3 m_3}-3\tilde a_{l_1 m_1}
\langle\tilde a_{l_2 m_2}
\tilde a_{l_3 m_3}\rangle_0
\right], 
\label{eq:cubic}
\end{equation}
where an angle bracket with subscript $0$ indicates an ensemble average 
over Gaussian simulations.
$\tilde a_{lm}$ is a harmonic coefficient obtained from observed (or simulated) data
maps with suitable weighting, which will be discussed in Section \ref{sec:weight}.
Eq. \eqref{eq:estim2} can be schematically represented as 
$\hat f^A_{\rm NL}=\displaystyle\sum_B
\langle S^A_{\rm prim}\rangle_{f^{B}_{\rm NL}=1}^{-1}
S^B_{\rm prim}$.
In our analysis, to estimate $\langle \tilde a_{lm}\tilde a_{l'm'}\rangle_0$
we accumulated at least $250$ Monte Carlo (MC) samples.

Assuming the Gaussianity for $X^A$, we estimate 
the covariance of the estimator $\hat f^A_{\rm NL}$, 
\begin{eqnarray}
\langle \hat f^A_{\rm NL}\hat f^B_{\rm NL}\rangle_0&=&\sum_{A'B'}
\langle S^A_{\rm prim}\rangle_{f^{A'}_{\rm NL}=1}^{-1}
\langle S^{A'}_{\rm prim}S^{B'}_{\rm prim}\rangle_0
\langle S^B_{\rm prim}\rangle_{f^{B'}_{\rm NL}=1}^{-1}\notag \\
&=&\langle S^A_{\rm prim}\rangle_{f^{B}_{\rm NL}=1}^{-1}. 
\end{eqnarray}
Here we used the relation 
$\langle S^A_{\rm prim}\rangle_{f^B_{\rm NL}=1}=
\langle S^A_{\rm prim}S^B_{\rm prim}\rangle_0$. 
See Appendix \ref{sec:equiv} for the derivation.

The estimator in Eq.~\eqref{eq:estim2} can be regarded as the generalization of a 
fast estimator in Ref. \cite{Komatsu:2003iq} to the case where
the initial perturbations are mixture of non-Gaussian adiabatic and 
uncorrelated isocurvature perturbations.
When we assume that either of adiabatic or uncorrelated isocurvature perturbations
are non-Gaussian and the other is Gaussian, Eq.~\eqref{eq:estim2}
can be reduced to
\begin{equation}
\hat f^A_{\rm NL}=S_{\rm prim}^A/\langle S_{\rm prim}^A\rangle_{f^A_{\rm NL}=1},
\end{equation}
where the subscripts $A$ indicates the perturbations which are assumed to be 
non-Gaussian. In this case, the variance of $\hat f^A_{\rm NL}$ is given by 
$1/\langle S^A_{\rm prim}\rangle_{f^A_{\rm NL}=1}$.

We note that there is a difficulty in computing the normalization factor 
$\langle S^A_{\rm prim}\rangle_{f^B_{\rm NL}=1}$. If we naively evaluate 
Eq. \eqref{eq:cubic} 
by taking ensemble average over simulated non-Gaussian CMB maps with 
small but non-zero $f_{\rm NL}^A$, 
a substantial number of simulations are required due to a large Gaussian fluctuation.
Instead, we divide $\tilde a_{lm}$ into its Gaussian and non-Gaussian
parts and evaluate $S^A_{\rm prim}$ without terms which are to vanish
by averaging. The details are presented in Appendix \ref{sec:est}.
This treatment also removes the need for setting a non-zero fiducial $\alpha$.

\subsection{Non-Gaussian CMB simulation}
\label{sec:sim}

In order to determine the normalization $\langle S^A_{\rm prim}\rangle_{f^B_{\rm NL}=1}$, we 
need to simulate non-Gaussian CMB maps.
With the Fourier transformation, Eq. \eqref{eq:alm} can be rewritten as 
\begin{equation}
a^A_{lm}=\int dr r^2 \alpha^A_l(r)\int d\hat r Y^*_{lm}(\hat r)X^A(\vec r).
\label{eq:almreal}
\end{equation}
In the case of local-type non-Gaussianity, 
Eq. \eqref{eq:almreal} allows us to simulate the non-Gaussian CMB maps
exactly \cite{Liguori:2003mb,Liguori:2007sj}.
We adopt the fast method developed in Ref. \cite{Elsner:2009md}, which
enables simulations of the local-type non-Gaussianity with sufficient speed.
The key technique in the method of Ref. \cite{Elsner:2009md} is that
the line of sight integral in Eq. \eqref{eq:almreal} is evaluated by the Gaussian quadrature
with optimized nodes and weights.
In our simulation, we optimized the nodes so that the mean square of the
error in simulated maps $a_{lm}$ at each multipole $(lm)$ should be less than $0.01$.
For $l_{\rm max}=1024$, we found that this level of accuracy requires 42 and 15 
nodes for curvature and isocurvature perturbations, respectively.

\subsection{Optimally-weighted CMB maps}
\label{sec:weight}

In estimating $\hat f^A_{\rm NL}$, 
we need to suitably weight observed (and simulated) maps 
$a_{lm}$ to obtain $\tilde a_{lm}$, 
in order to take into account the sensitivity and resolution of the survey.
As discussed in Ref. \cite{Smith:2009jr}, the optimal weight
is the inverse of the variance. This can be schematically represented as 
$\tilde a_{lm}=[C^{-1}a]_{lm}$,
where $C=C_S+C_N$ is the covariance matrix, with
those of signal and noise being denoted as $C_S$ and $C_N$, respectively.
While $C_S$ is diagonal in the harmonic space, 
$C_N$ is in general complicated for real observations.
In particular, WMAP and many CMB surveys have multiple channels, 
so that we need to take into account different beam functions, 
inhomogeneous noises, and a sky cut. In such a case, $\tilde a_{lm}$ 
can be given in an implicit form as
\begin{equation}
(C_S^{-1}+C_N^{-1})C_S\tilde a=C_N^{-1}a, 
\label{eq:atilde}
\end{equation}
where $C_N^{-1}=\displaystyle\sum_i b^{(i)}{C_N^{(i)}}^{-1}b^{(i)}$ and 
$C_N^{-1}a=\displaystyle\sum_i {C_N^{(i)}}^{-1}a^{(i)}$, with the subscript 
$i$ indicating a channel of the survey.
$b^{(i)}$ here is the beam function
for channel $i$.

Since the matrix $(C_S^{-1}+C_N^{-1})$ is 
dense in both the harmonic and pixel spaces, 
direct implementation of inversion $(C_S^{-1}+C_N^{-1})^{-1}$
is substantially prohibited. 
Instead, Eq. \eqref{eq:atilde} 
can be solved by the conjugate gradient (CG) method with 
preconditioning \cite{Oh:1998sr}.
How fast a CG method converges crucially depends on a
choice of pre-conditioner.
We adopt a fast method developed in Ref. \cite{Smith:2007rg}, 
which makes use of the multi-grid pre-conditioning.
This method also allows to marginalize over 
amplitude of components whose spatial templates $\{\tau\}$ are provided.
To do this, $C_N^{-1}$ is replaced
with $C_N^{-1}-\displaystyle\sum_{ab}C_N^{-1}\tau_a(\tau_aC_N^{-1}\tau_b)^{-1}C_N^{-1}\tau_b$, 
where the subscripts $a$ and $b$ indicate template components.
For the details of the method, we refer to Refs. 
\cite{Smith:2007rg,Smith:2009jr}.

\subsection{Data and assumption on cosmological model}
\label{sec:data}

In our analysis, we assume a flat power-law $\Lambda$CDM model as a
fiducial one and adopt the mean values for the cosmological parameters from the 
WMAP 7-year data alone \cite{Komatsu:2010fb}, 
\begin{equation}
(\omega_b,\,\omega_c,\,h,\,\tau,\,n_s,\,A_s)=(0.02249,\, 0.112,\,0.727,\,0.088,\,0.967,\,2.43\times10^{-9}),
\label{eq:cparams}
\end{equation}
where $\omega_b=\Omega_bh^2$ and $\omega_c=\Omega_ch^2$ are respectively the density parameters for baryon and cold dark matter, 
$h$ is the Hubble constant in units of 100km/s/Mpc, $\tau$ is the optical depth of reionization, and $n_s$ and $A_s$ 
are respectively the spectral index and amplitude of the power spectrum of curvature perturbations at a reference scale 
$k_*=0.002{\rm Mpc}^{-1}$, i.e., $P_\zeta(k)=\frac{2\pi^2}{k^3}A_s\left(\frac k{k_*}\right)^{n_s-1}$.

We also assume that the power spectrum of isocurvature perturbations
is in a power-law form $P_{S_{\rm G}}(k)\propto k^{n_{\rm iso}-4}$, and 
with regard to the fiducial value of the spectral index $n_{\rm iso}$ we
consider two different cases, $n_{\rm iso}=0.963(=n_{\rm adi})$ and $n_{\rm iso}=1$.
The transfer function of CMB is computed using 
the CAMB code \cite{Lewis:1999bs}.

We combine the foreground-cleaned maps of V and W bands of the WMAP 7-year data 
\cite{Jarosik:2010iu,Gold:2010fm}\footnote{http://lambda.gsfc.nasa.gov}
with a resolution $N_{\rm side}=512$ 
of the HEALPix pixelization scheme \cite{Gorski:2004by}
\footnote{http://healpix.jpl.nasa.gov}. We adopt the KQ75y7 mask \cite{Gold:2010fm}, 
which cuts 28.4 \% of the sky. We also set the maximum multipole 
$l_{\rm max}$ to 1024 in our analysis. 
We marginalize the amplitudes of the monopole $l=0$ and dipoles $l=1$ as default.
We also optionally marginalize the amplitudes of 
Galactic foreground components at large angular scales using 
the templates for the synchrotron, free-free and dust emissions from Ref. \cite{Gold:2010fm}.

\section{Result}
\label{sec:result}

Our results of constraints on the non-linearity parameters are summarized in Table \ref{tbl:const}.

\begin{table}[htb]
  \begin{center}
  \begin{tabular}{llcc}
  \hline
  \hline
  \multicolumn{2}{c}{setups}& $f_{\rm NL}$ & $\alpha^2f_{\rm NL}^{\rm (ISO)}$ \\
  \hline
  $n_{\rm iso}=0.963$
  &w/o template marginalization & $31\pm21$ &$5\pm63$ \\
  && $(36\pm23)$ & $(-39\pm 69)$\\
  \cline{2-4} 
  &w/ template marginalization & $32\pm21$ & $40\pm63$ \\
  && $(32\pm23)$ & $(0\pm70)$ \\
  \hline
  $n_{\rm iso}=1$ 
  &w/o template marginalization & $31\pm21$ & $19\pm62$ \\
  && $(34\pm23)$ &  $(-22\pm70)$ \\
  \cline{2-4} 
  &w/ template marginalization & $32\pm21$ & $40\pm66$\\
  && $(38\pm24)$ & $(-8\pm72)$ \\
  \hline
  \hline 
\end{tabular}
  \caption{
  Constraints on $f_{\rm NL}$ and $\alpha^2f_{\rm NL}^{\rm (ISO)}$ for different setups.
  We adopted four different setups regarding the fiducial value of $n_{\rm iso}$ and 
  template marginalization of the Galactic foregrounds. Constraints without parenthesis
  are estimated by fixing the other non-linearity parameter to zero.
  On the other hand, ones with parenthesis are estimated by marginalizing 
  over the other non-linearity parameter.
  }
  \label{tbl:const}
\end{center}
\end{table}

To check our analysis method, we evaluate
constraints on the non-linearity parameter 
for adiabatic perturbations, $f_{\rm NL}$, assuming isocurvature perturbations being absent, 
and compare them with those  in a previous study.
We obtain $f_{\rm NL}=31\pm21$ at 1$\sigma$ level
without template marginalization of the Galactic foregrounds. 
The central value 
by about 0.5$\sigma$ deviates from that of Ref. \cite{Komatsu:2010fb}, which
gives $f_{\rm NL}=42\pm21$ from the foreground-cleaned V+W band data 
with a resolution $N_{\rm side}=1024$. 
Since there are substantial differences between our analysis and that of Ref. 
\cite{Komatsu:2010fb}, such as  fiducial cosmological models 
and resolutions of the maps used,
we believe that this level of difference is acceptable and our result
is consistent with the previous study.
With template marginalization, we obtain $f_{\rm NL}=32\pm21$, 
which is exactly the same as the one given  by the WMAP group \cite{Komatsu:2010fb}.
We found that  template marginalization little affects constraints on $f_{\rm NL}$.

Now we present constraints on non-Gaussianity in uncorrelated isocurvature 
perturbations. 
We first assume that the adiabatic perturbations are Gaussian and fix $f_{\rm NL}$ to be zero.
For the cases of $n_{\rm iso}=0.963$ and $n_{\rm iso}=1$, 
we respectively obtain 
$\alpha^2 f_{\rm NL}^{\rm (ISO)}=5\pm63$ and $19\pm62$ at 1$\sigma$ level
without template marginalization. 
With template marginalization, these change to $40\pm63$ and $40\pm66$.
We found that the constraints on $\alpha^2f_{\rm NL}^{\rm (ISO)}$ are not strongly affected by
the fiducial value of $n_{\rm iso}$.
On the other hand, the constraints are more or less dependent on the treatment of the Galactic
foregrounds. The central values can change by 0.5$\sigma$ while 
the error remains almost unchanged. However, this shows that the effects
of residual foregrounds are not severe.

\begin{figure}
  \begin{center}
    \hspace{0mm}\scalebox{1.0}{\includegraphics{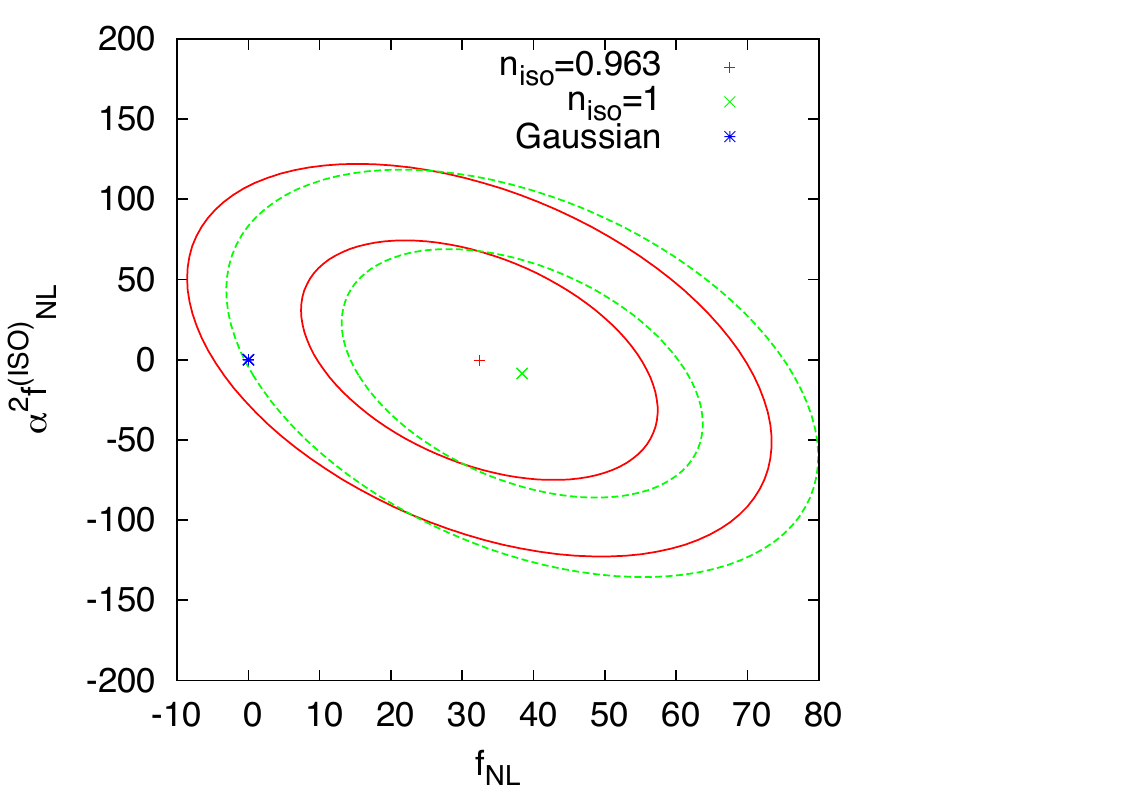}}
  \end{center}
  \caption{
  Joint constraints on $f_{\rm NL}$ and 
  $\alpha^2f_{\rm NL}^{\rm (ISO)}$ at 1 and 2$\sigma$ levels with template marginalization.
  Solid red and dashed green lines respectively show the cases for $n_{\rm iso}=0.963$ and $n_{\rm iso}=1$, 
  with the central values being indicated by the red plus and green cross.
  Blue star corresponds to the case where the perturbations are purely Gaussian.
  }
  \label{fig:contour}
\end{figure}

When we assume that both adiabatic and isocurvature perturbations
can be non-Gaussian,  we obtain a joint constraint on $(f_{\rm NL},\alpha^2f_{\rm NL}^{\rm (ISO)})$.
For the cases of $n_{\rm iso}=0.963$ and $n_{\rm iso}=1$, 
we respectively obtain $(f_{\rm NL},\alpha^2f_{\rm NL}^{\rm (ISO)})=(36\pm23,-39\pm69)$
and $(34\pm23,-22\pm70)$ without template marginalization.
With template marginalization, these changes to 
$(f_{\rm NL},\alpha^2f_{\rm NL}^{\rm (ISO)})=(32\pm23,0\pm70)$ and $(38\pm24,-8\pm72)$.
The error for each non-linearity parameter here is estimated
by marginalizing over the other non-linearity parameter.
In Fig.~\ref{fig:contour}, we show 2D constraints in the 
$f_{\rm NL}$-$\alpha^2f_{\rm NL}^{\rm (ISO)}$ plane for the cases 
with template marginalization.
Due to the correlation of 
$\alpha^2f^{\rm (ISO)}_{\rm NL}$ with $f_{\rm NL}$, a simultaneous 
fit for both of these variables changes the central value of $\alpha^2f^{\rm (ISO)}_{\rm NL}$, which
are not so constrained as $f_{\rm NL}$. 
We found that the dependence on $n_{\rm iso}$ is weak and 
the central values of $\alpha^2f_{\rm NL}$ can change by 0.5$\sigma$
by the treatment of the Galactic foregrounds.

In our analysis, we omitted effects of unresolved point sources, which
may bias our constraints on $f_{\rm NL}$ and $\alpha^2 f_{\rm NL}^{\rm (ISO)}$.
For the case of purely adiabatic perturbations, 
Ref. \cite{Komatsu:2010fb} studies effects of unresolved point sources and concludes that 
$f_{\rm NL}$ can be biased by 2 when the WMAP 7-year data is used. 
Because effects of unresolved point sources should be quite small at large angular scales $l<\mathcal O(100)$, 
where uncorrelated isocurvature perturbations can be prominent, we expect bias in 
$\alpha^2f_{\rm NL}^{\rm (ISO)}$ induced by unresolved point sources should be even smaller.  
We thus conclude that our constraints should be little affected by unresolved point sources.

As stated in Section \ref{sec:data}, 
our constraints are derived with other cosmological parameters being fixed. 
However, these parameters themselves have uncertainties, which can bias and/or weaken our constraints. 
Following the method of Ref. \cite{Liguori:2008vf}, we here discuss size of errors on the nonlinearity 
parameters coming from uncertainties in cosmological parameters. 
According to the study, given a difference in a cosmological parameters $\Delta p_i$, 
bias in a nonlinearity parameter $\Delta f_{\rm NL}^A$ can be approximately given by\footnote{ 
Eq. \eqref{eq:df} is not exactly the same as Ref. \cite{Liguori:2008vf}, in which 
only derivatives of bispectra, $\partial \hat b^A_{l_1l_2l_3}/\partial p_i$ are taken into account,
but those of the covariance matrix or power spectrum $\partial C_l/\partial p_i$ should not.
However, since there should not be severe cancelation between 
$\partial \hat b^A_{l_1l_2l_3}/\partial p_i$ and  $\partial C_l/\partial p_i$
for any of cosmological parameters we consider here, 
we believe Eq. \eqref{eq:df} should give rough estimates of bias in the non-linearity parameters
}
\begin{equation}
\Delta f_{\rm NL}^A \approx \sum_{BC} [F^{-1}]_{AB} 
\left[\frac{\partial F}{\partial p_i}\right]_{BC} f_{\rm NL}^C\Delta p_i, 
\label{eq:df}
\end{equation}
where $F_{AB}$ is the Fisher matrix for non-linearity parameters, 
\begin{equation}
F_{AB}=\frac16\sum_{l_1l_2l_3}\frac{(2l_1+1)(2l_2+1)(2l_3+1)}{4\pi}
\begin{pmatrix}
l_1 & l_2 & l_3 \\
0 & 0 & 0 
\end{pmatrix}^2
\frac{\hat b_{l_1l_2l_3}^A\hat b_{l_1l_2l_3}^B}{C_{l_1}C_{l_3}C_{l_2}}.
\end{equation}
We estimate $\Delta f_{\rm NL}^A$ for each of the cosmological parameters, 
whose uncertainties $\Delta p_i$ are taken from the constraint from the WMAP 7-year 
data alone in Ref. \cite{Komatsu:2010fb}. 
For the fiducial values of nonlinearity parameters, we here consider two extreme cases, 
$(f_{\rm NL}, \alpha^2f_{\rm NL}^{\rm (ISO)})=(70,\,0)$ and $(0,\,70)$, which are 
at 1 $\sigma$ deviation with current constraints.
For the case of $(f_{\rm NL}, \alpha^2f_{\rm NL}^{\rm (ISO)})=(70,\,0)$, 
$\Delta f_{\rm NL}$ can be as large as 5 from uncertainties in $n_s$ and $A_s$, while 
$\Delta (\alpha^2f_{\rm NL}^{\rm (ISO)})$ can be as large as 8 from those in $\omega_c$ and $h$.
On the other hand, for the case of $(f_{\rm NL}, \alpha^2f_{\rm NL}^{\rm (ISO)})=(0,\,70)$, 
$\Delta f_{\rm NL}$ can be as large as 0.3 from uncertainties in $\omega_c$ and $n_s$, while 
$\Delta (\alpha^2f_{\rm NL}^{\rm (ISO)})$ can be as large as 7 from that in $\omega_c$.
From these estimates, we conclude that uncertainties in the cosmological parameters can
affect the limits on $f_{\rm NL}$ and $\alpha^2 f_{\rm NL}^{\rm (ISO)}$ by about twenty and ten percents, respectively.

Having all these results, we conclude that CMB data is consistent
with Gaussianity at 2$\sigma$ level, even if the uncorrelated 
CDM isocurvature perturbations are included. 

Let us compare our results with other studies. 
While our method is optimal based on the bispectrum, 
constraints on the same type of non-Gaussianity is 
studied in Ref. \cite{Hikage:2008sk} based on the Minkowski functionals, which gives
$\alpha^2 f^{\rm (ISO)}_{\rm NL}=-15\pm 60$, when
the adiabatic perturbations are assumed to be Gaussian and $n_{\rm iso}=1$.
We found our constraint is consistent with the previous study.
On the other hand, we cannot find any improvement in the constraint, 
although our method is based on the optimal estimator 
and should be better than suboptimal ones. This suggests that
the Minkowski functional method is almost optimal for
uncorrelated CDM isocurvature perturbations.
This can also be expected from the Fisher matrix analysis in Ref. \cite{Hikage:2009rt}, 
which gives a Cram\'er-Rao bound $\Delta (\alpha^2 f_{\rm NL})=60$.

As a joint constraint on $f_{\rm NL}$ and $f^{\rm (ISO)}_{\rm NL}$,
our results are the first one obtained from observed data. 
On the other hand, the same constraint 
is forecasted using the Fisher matrix analysis in Ref.~\cite{Hikage:2009rt}.
We note that 1$\sigma$ errors of our results are consistent with the forecast.

\section{Application to the axion model} \label{sec:axion}

In this section, we apply the constraints on the isocurvature non-Gaussianity 
to the axion isocurvature model. 
First, we shortly describe how non-Gaussian isocurvature perturbations
in axion CDM arise in the inflationary Universe based on the $\delta N$-formalism
\cite{Sasaki:1995aw,Lyth:2004gb}.
The baseline of our derivation is  the same as in Ref. \cite{Kawasaki:2008sn}
(See also Ref. \cite{Hikage:2008sk}).
Throughout this section, we denote the energy density of a component $i$ on the 
uniform density hyper-surface of the total matter with $\rho_i(\vec x)$.
We adopt the following non-linear definition of isocurvature perturbations in 
a component $i$, 
\begin{equation}
S_i(\vec x)=3(\zeta_i-\zeta)(\vec x),
\end{equation}
where $\zeta_i$ and $\zeta$ are respectively the curvature perturbations on the uniform density 
hyper-surfaces of the component $i$ and the total matter.
According to the $\delta N$-formalism, $\rho_i(\vec x)$ should be given by $\rho_i(\vec x)
=\bar\rho_ie^{(1+w_i)S_i(\vec x)}$, where $\bar\rho_i$ is the mean of $\rho_i(\vec x)$
and $w_i$ is the equation of state of the component $i$.

The axion is a pseudo Nambu-Goldstone boson of the Peccei-Quinn (PQ) 
U(1) symmetry, which
solves the strong CP problem in quantum chromodynamics (QCD). 
If the PQ symmetry is broken during inflation, the axion has a classical field value
$a_i=F_a\theta$ and a vacuum fluctuation $\delta a$, where $F_a$
is the axion decay constant and $\theta\in[-\pi,\pi]$ is the initial misalignment angle
during inflation.
At high temperature, the axion is massless. As temperature $T$
decreases the QCD phase transition takes place. 
At this moment, the axion becomes massive
and starts the coherent oscillation around the true vacuum.
This oscillation of the axion contributes to the energy density of CDM.
We assume that CDM in the Universe is a mixture of the axion
and other CDM components which are adiabatic.

The coherent oscillation of the axion synchronously 
starts on the uniform energy density hyper-surface of the total matter
at around $m_{\rm axion}(T)\simeq3H(T)$
\footnote{Here we assume that there are no isocurvature perturbations 
in neutrinos or, if any, extra radiations \cite{Kawasaki:2011rc}. Otherwise, the coherent oscillation
does not start synchronously on the uniform density hyper-surface and as a consequence, 
additional isocurvature perturbations in the axion can be induced.}, 
where $m_{\rm axion}$ and $H$ is the mass of axion and Hubble expansion rate, respectively.
The energy density of the coherent oscillation is proportional to the square
of its initial amplitude, $\rho_{\rm axion}(\vec x)\propto (a_i+\delta a(\vec x))^2$, 
which leads
\begin{equation}
e^{S_{\rm axion}(\vec x)}=1+2\frac{a_i\delta a(\vec x)}{a_*^2}+
\frac{\delta a^2(\vec x)-\langle \delta a^2\rangle}{a_*^2}, 
\end{equation}
where 
\begin{equation}
a_*^2\equiv a_i^2+\langle \delta a^2\rangle.
\end{equation}
Since other CDM components are assumed to be adiabatic, their energy density is uniform
on the uniform density hyper-surface of the total matter.
Therefore $\rho_{\rm CDM}(\vec x)$ should be given by
\begin{equation}
\rho_{\rm CDM}(\vec x)=\bar\rho_{\rm CDM}\left[(1-r)+re^{S_{\rm axion}}\right], 
\end{equation}
where $r=\bar\rho_{\rm axion}/\bar\rho_{\rm CDM}$
is the energy fraction of the axion in CDM and $(1-r)$ is that of other CDM components.
Thus we finally obtain the isocurvature perturbations in CDM, 
\begin{eqnarray}
S_{\rm CDM}=\ln\left[\frac{\rho_{\rm CDM}(\vec x)}{\bar\rho_{\rm CDM}}\right]&
=&\ln\left[1-r+r e^{S_{\rm axion}}\right] \notag\\
&\simeq&2r\frac{a_i\delta a}{a_*^2}+\left[\frac1{4r}\frac{a_*^2}{a_i^2}\right]
\left[\left(2r\frac{a_i\delta a}{a_*^2}\right)^2-\Bigg\langle
\left(2r\frac{a_i\delta a}{a_*^2}\right)^2\Bigg\rangle\right]
+\cdots.
\end{eqnarray}
The last equality is approximately valid for $r\ll1$.
In the following, we keep  terms up to the second order.

According to Refs. \cite{Turner:1985si,Kawasaki:2008sn}, 
$r$ is given by 
\begin{equation}
r=0.2\omega_c^{-1}\left(\frac{F_a}{10^{12}\mbox{GeV}}\right)^{-0.82}
\left[\left(\frac{F_a}{10^{12}\mbox{GeV}}\right)^2\theta^2
+\left(\frac{H_{\rm inf}/2\pi}{10^{12}\mbox{GeV}}\right)^2\right], 
\end{equation}
where the first and second terms in the square bracket correspond to
contributions from the classical field $a_i$ and the fluctuation $\delta a$, respectively.
The fluctuation of axion $\delta a$ is almost scale-invariant (See also Ref. \cite{Kasuya:2009up}) 
and its root mean square is given by $\sqrt{\langle \delta a^2\rangle}=H_{\rm inf}/2\pi$, 
where $H_{\rm inf}$ is the Hubble scale during inflation.
Then power spectrum and bispectrum of CDM isocurvature perturbations
are given by\footnote{
Although, in the analysis in the previous sections, 
we omitted the loop contributions, we include them here. 
We also assume that logarithmic factors appear in the expression 
are considered to be $\mathcal{O}(1)$ and we here set them to be unity.  
}
\begin{eqnarray}
\langle S_{\rm CDM}(\vec k_1)S_{\rm CDM}(\vec k_2)\rangle
&=&4r^2\frac{(H_{\rm inf}/2\pi)^2}
{(F_a\theta)^2+(H_{\rm inf}/2\pi)^2}\frac{2\pi^2}{k_1^3}
(2\pi)^3\delta^{(3)}(\vec k_1+\vec k_2), \\
\langle S_{\rm CDM}(\vec k_1)S_{\rm CDM}(\vec k_2) 
S_{\rm CDM}(\vec k_3)\rangle 
&=&8r^3
\left[\frac{(H_{\rm inf}/2\pi)^2}{(F_a\theta)^2+(H_{\rm inf}/2\pi)^2}\right]^2 \\
&\times&\left[
\frac{(2\pi^2)^2}{k_1^3k_2^3}
+(\mbox{2 perms})
\right]
(2\pi)^3\delta^{(3)}(\vec k_1+\vec k_2+\vec k_3). \notag
\end{eqnarray}

Having all these results, 
constraints should be 
\begin{eqnarray}
0.2\left(\frac{F_a}{10^{12}\mbox{GeV}}\right)^{-0.82}
\left[\left(\frac{F_a}{10^{12}\mbox{GeV}}\right)^2\theta^2
+\left(\frac{H_{\rm inf}/2\pi}{10^{12}\mbox{GeV}}\right)^2\right]&<&\omega_c, 
\label{eq:axion1}
\\
0.16\left(\frac{F_a}{10^{12}\mbox{GeV}}\right)^{-1.64}
\left[\left(\frac{F_a}{10^{12}\mbox{GeV}}\right)^2\theta^2
+\left(\frac{H_{\rm inf}/2\pi}{10^{12}\mbox{GeV}}\right)^2\right]
\left(\frac{H_{\rm inf}/2\pi}{10^{12}\mbox{GeV}}\right)^2
&<&0.15 A_s\omega_c^2, 
\label{eq:axion2}\\
0.063
\left(\frac{F_a}{10^{12}\mbox{GeV}}\right)^{-2.46}
\left[\left(\frac{F_a}{10^{12}\mbox{GeV}}\right)^2\theta^2
+\left(\frac{H_{\rm inf}/2\pi}{10^{12}\mbox{GeV}}\right)^2\right]
\left(\frac{H_{\rm inf}/2\pi}{10^{12}\mbox{GeV}}\right)^4&<&140A_s^2\omega_c^3. 
\label{eq:axion3}
\end{eqnarray}
From top to bottom, these three equations correspond 
to the constraints on the abundance $r<1$, 
the isocurvature power spectrum $\alpha<0.15$ (95 \% CL) \cite{Komatsu:2010fb}, 
and non-Gaussianity $|\alpha^2f^{\rm (ISO)}_{\rm NL}|<140$ (2$\sigma$ level)
from our results, respectively. We should, however, note that 
since loop contributions are omitted in deriving the constraints in the previous sections,
thus our results cannot in a rigorous manner be applied 
to cases for $F_a\theta\lesssim H_{\rm inf}/2\pi$, 
where loop contributions dominate polyspectra.
Therefore, in the parameter regions with $F_a\theta\lesssim H_{\rm inf}/2\pi$, 
our constraints should be deemed as rough estimate.

\begin{figure}
  \begin{center}
  \begin{tabular}{cc}
    \hspace{0mm}\scalebox{.7}{\includegraphics{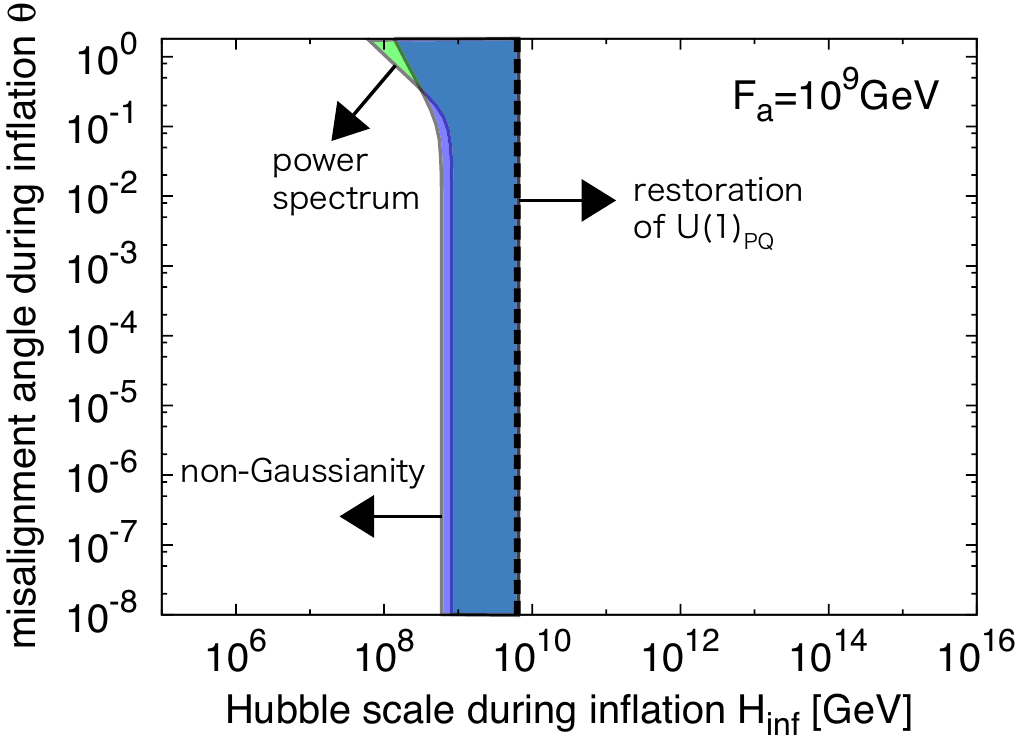}} &
    \hspace{0mm}\scalebox{.7}{\includegraphics{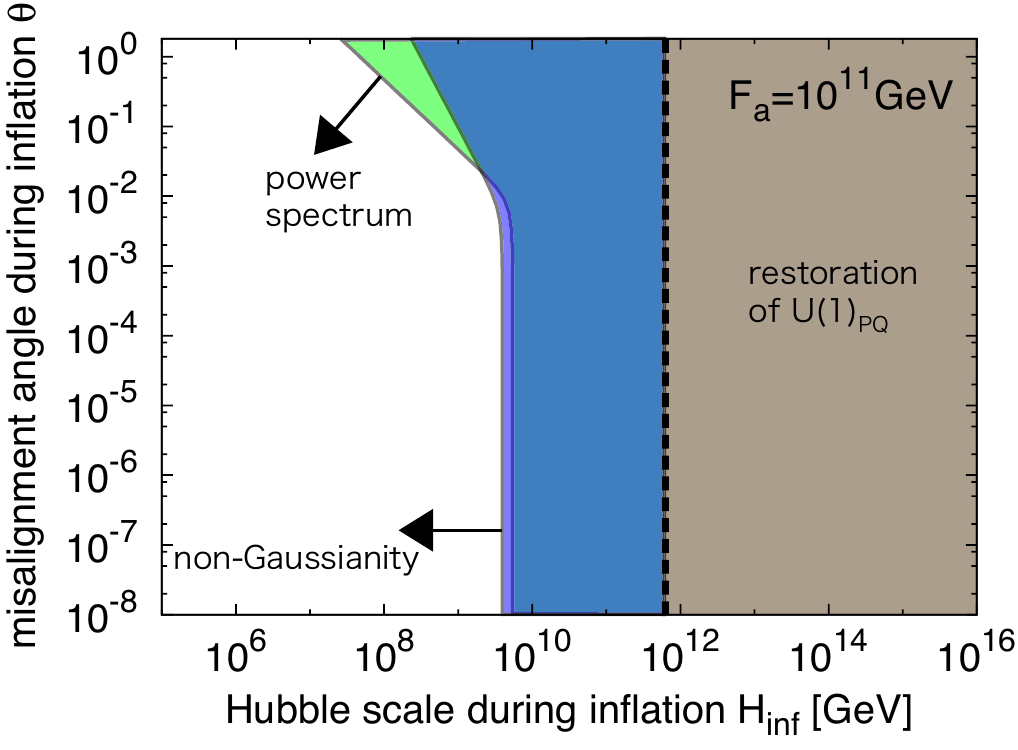}} \\
    \hspace{0mm}\scalebox{.7}{\includegraphics{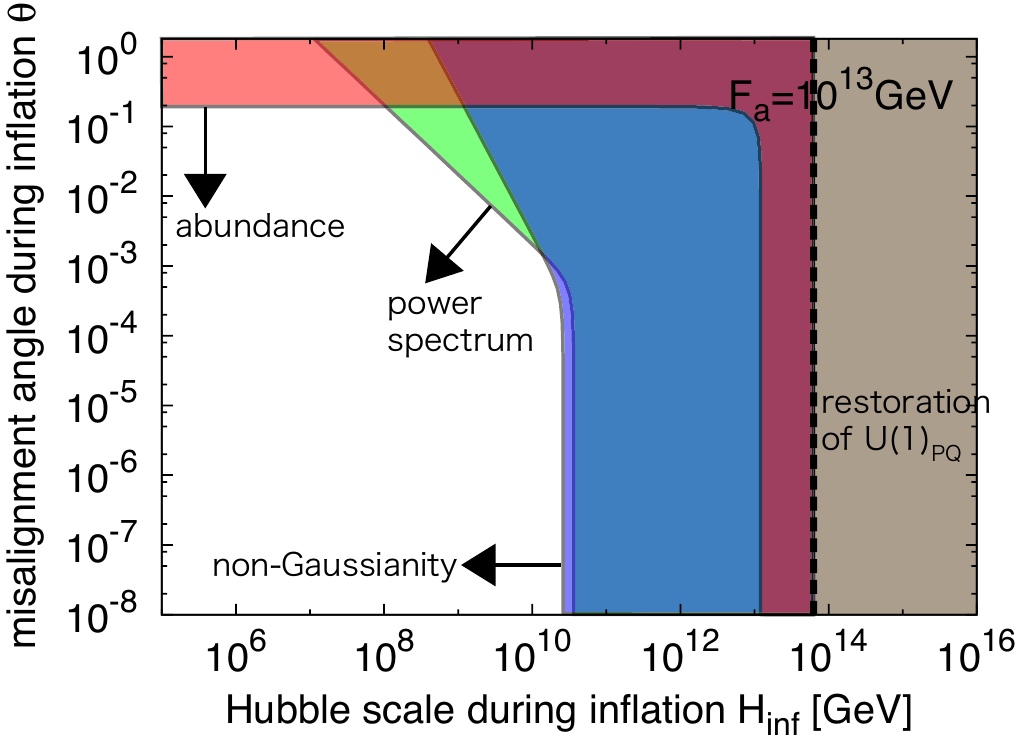}} &
    \hspace{0mm}\scalebox{.7}{\includegraphics{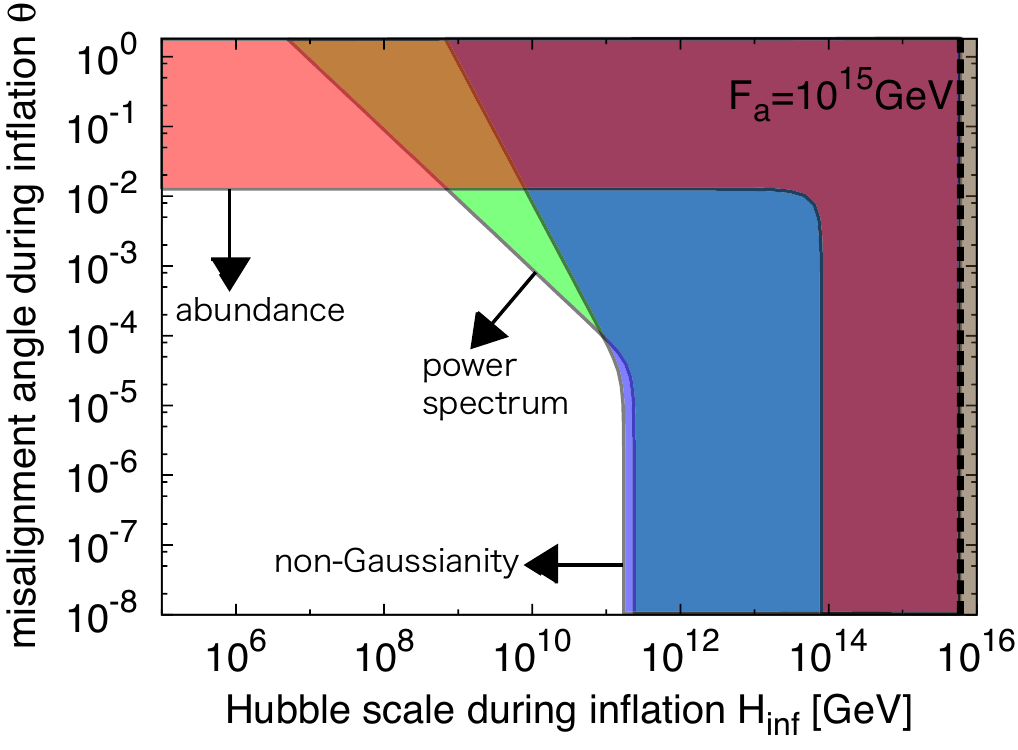}}
  \end{tabular}
  \end{center}
  \caption{
  Constraints on axion isocurvature model in the
  $H_{\rm inf}$-$\theta$ plane for several fixed values of $F_a$.
  Shaded regions are excluded by cosmological considerations. 
  At small $\theta$, the constraint on $H_{\rm inf}$ from the non-Gaussianity
  is slightly better than one from the power spectrum.
  See text for details.
  }
  \label{fig:H_theta}
\end{figure}

In Fig.~\ref{fig:H_theta}, we plotted the constraints of Eqs. \eqref{eq:axion1}-\eqref{eq:axion3}
in the $H_{\rm inf}$-$\theta$ plane with several fixed values of $F_a$, 
with $\omega_c$ and $A_s$ being fixed to the values given in Eq. \eqref{eq:cparams}.
Constraints have critical points at $F_a\theta=H_{\inf}/2\pi$, where
contributions of the classical field value and the fluctuation
are comparable. We found that the constraint on non-Gaussianity 
from our results gives an upper bound on $H_{\rm inf}$ comparable to that
from one on the isocurvature power spectrum. 

When $H_{\rm inf}/2\pi>F_a$, the PQ symmetry restores
during inflation\footnote{Here we assumed 
the reheating temperature $T_{\rm reh}$ is below $H_{\rm inf}$.
If $T_{\rm reh}>H_{\rm inf}$, 
the PQ symmetry can restore for smaller $H_{\rm inf}$ after inflation.}
and in that case CDM axions are produced
from the system of axionic string-wall system as well as the
coherent oscillation with an initial alignment angle $\theta=\mathcal O(1)$. 
This excludes $H_{\rm inf}/2\pi>F_a$ for $F_a \gtrsim 10^{11}$GeV in Fig. \ref{fig:H_theta}
(For details, we refer to e.g. Refs. \cite{Hiramatsu:2010yu,Hiramatsu:2012gg}). 

These constraints lead that
even if $\theta$ can be small by chance, 
large $F_a\gtrsim10^{11}$GeV is not allowed unless the inflation scale is 
low $H_{\rm inf}\lesssim10^{11}$GeV.

We note that in deriving the constraints above, we fixed the values of $\omega_c$ and $A_s$, which in principle 
have uncertainties themselves. Since these parameters are precisely determined to an accuracy of several percents
from current data \cite{Komatsu:2010fb}, their uncertainties can be safely omitted in the right hand sides of Eqs. 
\eqref{eq:axion1}-\eqref{eq:axion3}.
On the other hand, as is discussed in the previous section, more significant are the effects of uncertainties in the cosmological 
parameters on the estimation of $\alpha^2f_{\rm NL}^{\rm (ISO)}$, 
which can increase the right hand side of Eq. \eqref{eq:axion3} by about ten percents. 
Still, Fig. \ref{fig:H_theta} is not affected very much, because the bounds on $H_{\rm inf}$ and $\theta$ are 
proportional to some fractional powers of the right hand side of Eq. \eqref{eq:axion3}.

\section{Conclusion} \label{sec:conclusion}

We studied constraints on non-Gaussianity in a mixture of 
adiabatic and uncorrelated CDM isocurvature perturbations, 
which should be regarded as an extension of the adiabatic local-type one
to non-adiabatic primordial perturbations.
We adopted the optimal bispectrum estimator for the
non-linearity parameters, and a fast method for
non-Gaussian CMB simulation and the optimal weighting of 
observed maps are integrated in our analysis.
Using the WMAP 7-year data of CMB temperature maps with template marginalization
of the Galactic foregrounds, 
we obtained a constraint $\alpha^2 f_{\rm NL}^{\rm (ISO)}=40\pm62$
at 1$\sigma$ level for the scale invariant isocurvature power spectrum
when the adiabatic perturbations are assumed to be Gaussian.
Under the same setup, we also obtained a joint constraint on the non-linearity 
parameters $(f_{\rm NL}, \alpha^2 f^{\rm (ISO)}_{\rm NL})=(38\pm 24,-8\pm 72)$. 
The constraints weakly depend on the fiducial value of the isocurvature spectral index.
Effects of the Galactic foregrounds at large angular scales are not severe.
We found no statistically significant deviation from Gaussianity and 
the current WMAP observation of CMB is consistent with Gaussianity
even when we include this type of isocurvature perturbations.
We applied our results to the axion model.

Since the CDM and baryon isocurvature modes affect the CMB 
anisotropy in the same way except for the overall amplitude, 
our constraint can be translated into the baryon isocurvature perturbations.
This can be easily done by substituting 
$\left(\frac{\Omega_b}{\Omega_c}\right)^3 \alpha^2f^{\rm (ISO)}_{\rm NL}$
for $\alpha^2 f^{\rm (ISO)}_{\rm NL}$, where $\Omega_b$ and $\Omega_c$ 
are the density parameters of baryon and CDM, respectively.
Although we restrict ourselves to uncorrelated CDM isocurvature 
perturbations in this paper, our method can be generalized to the cases of 
non-Gaussianity in isocurvature perturbations correlated with adiabatic ones
and other types such as neutrino ones.
These will be studied in a forthcoming paper \cite{Hikage:2012tf}.

\bigskip
\bigskip

\noindent 
\section*{Acknowledgment}

The authors would like to thank Eiichiro Komatsu and Kazunori Nakayama 
for helpful discussion.
T.~S. would like to thank the Japan Society for the Promotion of Science for the financial report.
The authors acknowledge Kobayashi-Maskawa Institute for the Origin of
Particles and the Universe, Nagoya University for providing computing
resources  in conducting the research reported in this paper.
This work is supported by Grant-in-Aid for Scientific research from
the Ministry of Education, Science, Sports, and Culture (MEXT), Japan,
No.\ 14102004 (M.K.), No.\ 21111006 (M.K.), No.\ 23740195 (T.T.),  
No.\ 24740160 (C.H.) and also 
by World Premier International Research Center Initiative (WPI Initiative), MEXT, Japan. 
Some of the results in this paper have been derived using the HEALPix 
\cite{Gorski:2004by} package.
\appendix

\noindent 
\section{Equivalence between
$\langle S_{\rm prim}^A\rangle_{f^B_{\rm NL}=1}$ and 
$\langle S_{\rm prim}^AS_{\rm prim}^B\rangle_0$ }
\label{sec:equiv}

In this appendix, we show that the normalization factor
$\langle S_{\rm prim}^A\rangle_{f^B_{\rm NL}=1}$ and 
the covariance $\langle S_{\rm prim}^AS_{\rm prim}^B\rangle_0$ 
is equivalent.

From Eq. \eqref{eq:cubic}, we obtain
\begin{equation}
\langle S_{\rm prim}^A\rangle_{f^B_{\rm NL}=1}
=\frac16\sum_{\{lm\}}\hat b^A_{l_1l_2l_3}
\mathcal G^{l_1l_2l_3}_{m_1m_2m_3}
\langle\tilde a_{l_1m_1}
\tilde a_{l_2m_2}
\tilde a_{l_3m_3}
\rangle_{f^B_{\rm NL}=1}, 
\label{eq:app1}
\end{equation}
where we used the fact that $\langle \tilde a_{lm}\rangle_{f^A_{\rm NL}=1}=0$.
As discussed in Section \ref{sec:weight}, we take $\tilde a_{lm}=\displaystyle\sum_{l'm'}[C^{-1}]_{lm,l'm'}a_{l'm'}$, 
so that 
\begin{eqnarray}
\langle\tilde a_{l_1m_1}
\tilde a_{l_2m_2}
\tilde a_{l_3m_3}
\rangle_{f^B_{\rm NL}=1} &\\
&=& \sum_{\{l'm'\}}[C^{-1}]_{l_1m_1,l'_1m'_1}
[C^{-1}]_{l_2m_2,l'_2m'_2}[C^{-1}]_{l_3m_3,l'_3m'_3}
\langle a_{l'_1m'_1}a_{l'_2m'_2}a_{l'_3m'_3}\rangle_{f^B_{\rm NL}=1} \notag \\
&=& \sum_{\{l'm'\}}[C^{-1}]_{l_1m_1,l'_1m'_1}
[C^{-1}]_{l_2m_2,l'_2m'_2}[C^{-1}]_{l_3m_3,l'_3m'_3}
\mathcal G^{l'_1l'_2l'_3}_{m'_1m'_2m'_3}\hat b^B_{l'_1l'_2l'_3}. \notag
\label{eq:app2}
\end{eqnarray}
Combining Eqs. \eqref{eq:app1} and \eqref{eq:app2}, we obtain
\begin{equation}
\langle S_{\rm prim}^A\rangle_{f^B_{\rm NL}=1}=\frac16\sum_{\{lml'm'\}}
\hat b^A_{l_1l_2l_3}
\mathcal G^{l_1l_2l_3}_{m_1m_2m_3}[C^{-1}]_{l_1m_1,l'_1m'_1}
[C^{-1}]_{l_2m_2,l'_2m'_2}[C^{-1}]_{l_3m_3,l'_3m'_3}
\mathcal G^{l'_1l'_2l'_3}_{m'_1m'_2m'_3}\hat b^B_{l'_1l'_2l'_3}.
\label{eq:app3}
\end{equation}

On the other hand, 
\begin{eqnarray}
\langle S_{\rm prim}^AS_{\rm prim}^B\rangle_0
&=&\frac1{36}\sum_{\{lml'm'\}}\hat b^A_{l_1l_2l_3}\hat b^B_{l'_1l'_2l'_3}
\mathcal G^{l_1l_2l_3}_{m_1m_2m_3}
\mathcal G^{l'_1l'_2l'_3}_{m'_1m'_2m'_3}\left[
\langle\tilde a_{l_1m_1}
\tilde a_{l_2m_2}\tilde a_{l_3m_3}
\tilde a^*_{l'_1m'_1}\tilde a^*_{l'_2m'_2}
\tilde a^*_{l'_3m'_3}\rangle_0 \right.\\
&&\left.-6\langle
\tilde a_{l_1m_1}\tilde a^*_{l'_1m'_1}
\tilde a^*_{l'_2m'_2}\tilde a^*_{l'_3m'_3}\rangle_0
\langle\tilde a_{l_2m_2}\tilde a_{l_3m_3}\rangle_0
+9\langle \tilde a_{l_1m_1}\tilde a^*_{l'_1m'_1}\rangle_0
\langle\tilde a_{l_2m_2}\tilde a_{l_3m_3}\rangle_0
\langle\tilde a^*_{l'_2m'_2}\tilde a^*_{l'_3m'_3}\rangle_0\right]. \notag
\label{eq:app4}
\end{eqnarray}
By taking the Wick's contraction, the terms in the square bracket 
are reduced into
\begin{equation}
[C^{-1}]_{l_1m_1,l'_1m'_1}
[C^{-1}]_{l_2m_2,l'_2m'_2}
[C^{-1}]_{l_3m_3,l'_3m'_3}+\mbox{(5 perms)}, 
\label{eq:app5}
\end{equation}
where we used the fact that $\langle a_{lm}a_{l'm'}^*\rangle_0=C_{lm,l'm'}$.
Thus, we obtain
\begin{equation}
\langle S_{\rm prim}^AS_{\rm prim}^B\rangle_0=
\frac16\sum_{\{lml'm'\}}\hat b^A_{l_1l_2l_3}\hat b^B_{l'_1l'_2l'_3}
\mathcal G^{l_1l_2l_3}_{m_1m_2m_3}
\mathcal G^{l'_1l'_2l'_3}_{m'_1m'_2m'_3}
[C^{-1}]_{l_1m_1,l'_1m'_1}
[C^{-1}]_{l_2m_2,l'_2m'_2}[C^{-1}]_{l_3m_3,l'_3m'_3}.
\label{eq:app6}
\end{equation}

From Eqs. \eqref{eq:app3} and \eqref{eq:app6}, we 
finally see that $\langle S_{\rm prim}^A\rangle_{f^B_{\rm NL}=1}=
\langle S_{\rm prim}^AS_{\rm prim}^B\rangle_0$. 

\noindent 
\section{Estimation of normalization}
\label{sec:est}

Here we show how we can estimate the 
normalization factor $\langle S_{\rm prim}^A\rangle_{f^B_{\rm NL}=1}$
from simulation.

First we  divide the CMB anisotropy $a^A_{lm}$ into 
Gaussian and non-Gaussian parts 
\begin{equation}
a^A_{lm}=
a^A_{{\rm G},lm}+f^A_{\rm NL}a^A_{{\rm NG},lm}, 
\end{equation}
where $a^A_{{\rm G},lm}$ and $a^A_{{\rm NG},lm}$ respectively
are the Gaussian and non-Gaussian contributions, which are given by
\begin{eqnarray}
a^A_{{\rm G},lm}&=&\int dr r^2\alpha^A_l(r)\int d\hat r Y^*_{lm}(\hat r)X_{\rm G}^A(\vec r), \\
a^A_{{\rm NG},lm}&=&\int dr r^2\alpha^A_l(r)\int d\hat r Y^*_{lm}(\hat r)(X_{\rm G}^A(\vec r)^2
-\langle X_{\rm G}^A(\vec r)^2\rangle).
\end{eqnarray}
Since $\tilde a_{lm}$ is a linear combination of $a_{lm}$, 
we can also divide the weighted map into its Gaussian 
and non-Gaussian parts, $\tilde a^A_{lm}=\tilde a^A_{{\rm G},lm}+f^A_{\rm NL}\tilde a^A_{{\rm NG},lm}$.
Then $\langle S^A_{\rm prim}\rangle_{f^B_{\rm NL}=1}$ in Eq. \eqref{eq:app1} can be 
rewritten as 
\begin{eqnarray}
\langle S^A_{\rm prim}\rangle_{f^B_{\rm NL}=1}&=&
\frac16\sum_{\{lm\}}\hat b^A_{l_1l_2l_3}\mathcal G^{l_1l_2l_3}_{m_1m_2m_3}
\langle (\tilde a_{{\rm G},l_1m_1}+\tilde a^B_{{\rm NG},l_1m_1})
 (\tilde a_{{\rm G},l_2m_2}+\tilde a^B_{{\rm NG},l_2m_2})
 (\tilde a_{{\rm G},l_3m_3}+\tilde a^B_{{\rm NG},l_3m_3}) \rangle_{f^B_{\rm NL}=1} \notag\\
&=&
\frac16\sum_{\{lm\}}\hat b^A_{l_1l_2l_3}\mathcal G^{l_1l_2l_3}_{m_1m_2m_3}
\left(3
\langle\tilde a^B_{{\rm NG},l_1m_1}
\tilde a^B_{{\rm G},l_2m_2}\tilde a^B_{{\rm G},l_3m_3}\rangle
+\langle\tilde a^B_{{\rm NG},l_2m_2}\tilde a^B_{{\rm NG},l_2m_2}\tilde a^B_{{\rm NG},l_3m_3}\rangle\right)_{f^B_{\rm NL}=1}, \notag \\ 
\label{eq:appB}
\end{eqnarray}
where the last equality follows from the Wick's theorem. As long as non-Gaussianity
is not too large, the second term in the last line can be omitted.
To evaluate $\langle\tilde a^B_{{\rm NG},l_1m_1}
\tilde a^B_{{\rm G},l_2m_2}\tilde a^B_{{\rm G},l_3m_3}\rangle
$, we checked that $\mathcal O(10)$ MC samples are enough for 
convergence.
In addition, since the terms in the last line of Eq. \eqref{eq:appB} consists of 
only anisotropy $\tilde a^B_{lm}$ from a single perturbation mode $X^B$, 
we can estimate $\langle S^A_{\rm prim}\rangle_{f^B_{\rm NL}=1}$ without 
assuming a fiducial value for the fraction of the isocurvature contribution $\alpha$. 



\begin{thebibliography}{}

\bibitem{Komatsu:2010fb} 
  E.~Komatsu {\it et al.}  [WMAP Collaboration],
  Astrophys.\ J.\ Suppl.\  {\bf 192}, 18 (2011)
  [arXiv:1001.4538 [astro-ph.CO]].

\bibitem{Bucher:1999re} 
  M.~Bucher, K.~Moodley and N.~Turok,
  Phys.\ Rev.\ D {\bf 62}, 083508 (2000)
  [astro-ph/9904231].

\bibitem{Kawasaki:2011rc} 
  M.~Kawasaki, K.~Miyamoto, K.~Nakayama and T.~Sekiguchi,
  JCAP {\bf 1202}, 022 (2012)
  [arXiv:1107.4962 [astro-ph.CO]].
  

\bibitem{Kawakami:2012ke} 
  E.~Kawakami, M.~Kawasaki, K.~Miyamoto, K.~Nakayama and T.~Sekiguchi,
  JCAP {\bf 1207}, 037 (2012)
  [arXiv:1202.4890 [astro-ph.CO]].

\bibitem{Axenides:1983hj}
  M.~Axenides, R.~H.~Brandenberger, M.~S.~Turner,
  Phys.\ Lett.\  {\bf B126}, 178 (1983).
  
\bibitem{Seckel:1985tj}
  D.~Seckel, M.~S.~Turner,
  Phys.\ Rev.\  {\bf D32}, 3178 (1985).
  
\bibitem{Linde:1985yf}
  A.~D.~Linde,
  Phys.\ Lett.\  {\bf B158}, 375-380 (1985).
  
\bibitem{Linde:1990yj}
  A.~D.~Linde, D.~H.~Lyth,
  Phys.\ Lett.\  {\bf B246}, 353-358 (1990).
  
\bibitem{Turner:1990uz}
  M.~S.~Turner, F.~Wilczek,
  Phys.\ Rev.\ Lett.\  {\bf 66}, 5 (1991).
   
\bibitem{Linde:1991km}
  A.~D.~Linde,
  Phys.\ Lett.\  {\bf B259}, 38 (1991).
  
\bibitem{Lyth:1991ub}
  D.~H.~Lyth,
  Phys.\ Rev.\  {\bf D45}, 3394 (1992).
  
\bibitem{Kasuya:2009up}
  S.~Kasuya, M.~Kawasaki,
  Phys.\ Rev.\  {\bf D80}, 023516 (2009)
  [arXiv:0904.3800 [astro-ph.CO]].
  
\bibitem{Enqvist:1998pf}
  K.~Enqvist and J.~McDonald,
  Phys.\ Rev.\ Lett.\  {\bf 83}, 2510 (1999)
  [arXiv:hep-ph/9811412].
  
  
\bibitem{Enqvist:1999hv}
  K.~Enqvist, J.~McDonald,
  Phys.\ Rev.\  {\bf D62}, 043502 (2000)
  [hep-ph/9912478].
  
  
\bibitem{Kawasaki:2001in}
  M.~Kawasaki and F.~Takahashi,
  Phys.\ Lett.\  B {\bf 516}, 388 (2001)
  [arXiv:hep-ph/0105134].
  
\bibitem{Enqvist:2001zp}
K.~Enqvist and M.~S.~Sloth,
Nucl.\ Phys.\ B {\bf 626}, 395 (2002)
[arXiv:hep-ph/0109214].

\bibitem{Lyth:2001nq}
D.~H.~Lyth and D.~Wands,
Phys.\ Lett.\ B {\bf 524}, 5 (2002)
[arXiv:hep-ph/0110002].

\bibitem{Moroi:2001ct}
T.~Moroi and T.~Takahashi,
Phys.\ Lett.\ B {\bf 522}, 215 (2001)
[Erratum-ibid.\ B {\bf 539}, 303 (2002)]
[arXiv:hep-ph/0110096].





\bibitem{Moroi:2002rd}
  T.~Moroi and T.~Takahashi,
  Phys.\ Rev.\  D {\bf 66}, 063501 (2002)
  [arXiv:hep-ph/0206026].



\bibitem{Lyth:2002my}
  D.~H.~Lyth, C.~Ungarelli and D.~Wands,
  Phys.\ Rev.\  D {\bf 67}, 023503 (2003)
  [arXiv:astro-ph/0208055].

\bibitem{Moroi:2002vx} 
  T.~Moroi and H.~Murayama,
  Phys.\ Lett.\ B {\bf 553}, 126 (2003)
  [hep-ph/0211019].


\bibitem{Lyth:2003ip}
  D.~H.~Lyth and D.~Wands,
  Phys.\ Rev.\  D {\bf 68}, 103516 (2003)
  [arXiv:astro-ph/0306500].


\bibitem{Hamaguchi:2003dc} 
  K.~Hamaguchi, M.~Kawasaki, T.~Moroi and F.~Takahashi,
  Phys.\ Rev.\ D {\bf 69}, 063504 (2004)
  [hep-ph/0308174].

\bibitem{Gordon:2003hw} 
  C.~Gordon and K.~A.~Malik,
  Phys.\ Rev.\ D {\bf 69}, 063508 (2004)
  [astro-ph/0311102].

\bibitem{Ikegami:2004ve} 
  M.~Ikegami and T.~Moroi,
  Phys.\ Rev.\ D {\bf 70}, 083515 (2004)
  [hep-ph/0404253].

\bibitem{Ferrer:2004nv} 
  F.~Ferrer, S.~Rasanen and J.~Valiviita,
  JCAP {\bf 0410}, 010 (2004)
  [astro-ph/0407300].


\bibitem{Beltran:2008ei}
  M.~Beltran,
  Phys.\ Rev.\  D {\bf 78}, 023530 (2008)
  [arXiv:0804.1097 [astro-ph]].
   
\bibitem{Moroi:2008nn}
  T.~Moroi and T.~Takahashi,
  Phys.\ Lett.\  B {\bf 671}, 339 (2009)
  [arXiv:0810.0189 [hep-ph]].

\bibitem{Lemoine:2009yu} 
  M.~Lemoine, J.~Martin and J.~Yokoyama,
  Europhys.\ Lett.\  {\bf 89}, 29001 (2010)
  [arXiv:0903.5428 [astro-ph.CO]].

\bibitem{Lemoine:2009is} 
  M.~Lemoine, J.~Martin and J.~Yokoyama,
  Phys.\ Rev.\ D {\bf 80}, 123514 (2009)
  [arXiv:0904.0126 [astro-ph.CO]].

\bibitem{Takahashi:2009cx}
  T.~Takahashi, M.~Yamaguchi and S.~Yokoyama,
  Phys.\ Rev.\  D {\bf 80}, 063524 (2009)
  [arXiv:0907.3052 [astro-ph.CO]].
  
  
\bibitem{Kawasaki:2008sn} 
  M.~Kawasaki, K.~Nakayama, T.~Sekiguchi, T.~Suyama and F.~Takahashi,
  JCAP {\bf 0811}, 019 (2008)
  [arXiv:0808.0009 [astro-ph]].

\bibitem{Langlois:2008vk} 
  D.~Langlois, F.~Vernizzi and D.~Wands,
  JCAP {\bf 0812}, 004 (2008)
  [arXiv:0809.4646 [astro-ph]].

\bibitem{Kawasaki:2008pa} 
  M.~Kawasaki, K.~Nakayama, T.~Sekiguchi, T.~Suyama and F.~Takahashi,
  JCAP {\bf 0901}, 042 (2009)
  [arXiv:0810.0208 [astro-ph]].
  
\bibitem{Kawakami:2009iu} 
  E.~Kawakami, M.~Kawasaki, K.~Nakayama and F.~Takahashi,
  JCAP {\bf 0909}, 002 (2009)
  [arXiv:0905.1552 [astro-ph.CO]].

\bibitem{Langlois:2011zz} 
  D.~Langlois and A.~Lepidi,
  JCAP {\bf 1101}, 008 (2011)
  [arXiv:1007.5498 [astro-ph.CO]].


\bibitem{Langlois:2010fe} 
  D.~Langlois and T.~Takahashi,
  JCAP {\bf 1102}, 020 (2011)
  [arXiv:1012.4885 [astro-ph.CO]].
  
\bibitem{Moodley:2004nz} 
  K.~Moodley, M.~Bucher, J.~Dunkley, P.~G.~Ferreira and C.~Skordis,
  Phys.\ Rev.\ D {\bf 70}, 103520 (2004)
  [astro-ph/0407304].
  
\bibitem{Beltran:2004uv} 
  M.~Beltran, J.~Garcia-Bellido, J.~Lesgourgues and A.~Riazuelo,
  Phys.\ Rev.\ D {\bf 70}, 103530 (2004)
  [astro-ph/0409326].
  
\bibitem{Bean:2006qz} 
  R.~Bean, J.~Dunkley and E.~Pierpaoli,
  Phys.\ Rev.\ D {\bf 74}, 063503 (2006)
  [astro-ph/0606685].
  
\bibitem{Trotta:2006ww} 
  R.~Trotta,
  Mon.\ Not.\ Roy.\ Astron.\ Soc.\  {\bf 375}, L26 (2007)
  [astro-ph/0608116].
  
\bibitem{Keskitalo:2006qv} 
  R.~Keskitalo, H.~Kurki-Suonio, V.~Muhonen and J.~Valiviita,
  JCAP {\bf 0709}, 008 (2007)
  [astro-ph/0611917].
  
\bibitem{Kawasaki:2007mb} 
  M.~Kawasaki and T.~Sekiguchi,
  Prog.\ Theor.\ Phys.\  {\bf 120}, 995 (2008)
  [arXiv:0705.2853 [astro-ph]].
  
\bibitem{Sollom:2009vd} 
  I.~Sollom, A.~Challinor and M.~P.~Hobson,
  Phys.\ Rev.\ D {\bf 79}, 123521 (2009)
  [arXiv:0903.5257 [astro-ph.CO]].
  
  
\bibitem{Valiviita:2009bp} 
  J.~Valiviita and T.~Giannantonio,
  Phys.\ Rev.\ D {\bf 80}, 123516 (2009)
  [arXiv:0909.5190 [astro-ph.CO]].

\bibitem{Li:2010yb} 
  H.~Li, J.~Liu, J.~-Q.~Xia and Y.~-F.~Cai,
  Phys.\ Rev.\ D {\bf 83}, 123517 (2011)
  [arXiv:1012.2511 [astro-ph.CO]].

\bibitem{Valiviita:2012ub} 
  J.~Valiviita, M.~Savelainen, M.~Talvitie, H.~Kurki-Suonio and S.~Rusak,
  Astrophys.\ J.\  {\bf 753}, 151 (2012)
  [arXiv:1202.2852 [astro-ph.CO]].

\bibitem{Hikage:2008sk}
  C.~Hikage, K.~Koyama, T.~Matsubara, T.~Takahashi and M.~Yamaguchi,
  Mon.\ Not.\ Roy.\ Astron.\ Soc.\  {\bf 398}, 2188 (2009)
  [arXiv:0812.3500 [astro-ph]].

\bibitem{Hikage:2009rt} 
  C.~Hikage, D.~Munshi, A.~Heavens and P.~Coles,
  Mon.\ Not.\ Roy.\ Astron.\ Soc.\  {\bf 404}, 1505 (2010)
  [arXiv:0907.0261 [astro-ph.CO]].

\bibitem{Langlois:2011hn} 
  D.~Langlois and B.~van Tent,
  Class.\ Quant.\ Grav.\  {\bf 28}, 222001 (2011)
  [arXiv:1104.2567 [astro-ph.CO]].
  
\bibitem{Langlois:2012tm} 
  D.~Langlois and B.~van Tent,
  JCAP {\bf 1207}, 040 (2012)
  [arXiv:1204.5042 [astro-ph.CO]].

\bibitem{Hikage:2012tf} 
  C.~Hikage, M.~Kawasaki, T.~Sekiguchi and T.~Takahashi,
  arXiv:1212.6001 [astro-ph.CO].

\bibitem{Salopek:1990jq} 
  D.~S.~Salopek and J.~R.~Bond,
  Phys.\ Rev.\ D {\bf 42}, 3936 (1990).

\bibitem{Gangui:1993tt} 
  A.~Gangui, F.~Lucchin, S.~Matarrese and S.~Mollerach,
  Astrophys.\ J.\  {\bf 430}, 447 (1994)
  [astro-ph/9312033].

\bibitem{Komatsu:2001rj} 
  E.~Komatsu and D.~N.~Spergel,
  Phys.\ Rev.\ D {\bf 63}, 063002 (2001)
  [astro-ph/0005036].

\bibitem{Smith:2009jr} 
  K.~M.~Smith, L.~Senatore and M.~Zaldarriaga,
  JCAP {\bf 0909}, 006 (2009)
  [arXiv:0901.2572 [astro-ph]].

\bibitem{Komatsu:2003iq} 
  E.~Komatsu, D.~N.~Spergel and B.~D.~Wandelt,
  Astrophys.\ J.\  {\bf 634}, 14 (2005)
  [astro-ph/0305189].

\bibitem{Yadav:2007rk} 
  A.~P.~S.~Yadav, E.~Komatsu and B.~D.~Wandelt,
  Astrophys.\ J.\  {\bf 664}, 680 (2007)
  [astro-ph/0701921].

\bibitem{Creminelli:2005hu} 
  P.~Creminelli, A.~Nicolis, L.~Senatore, M.~Tegmark and M.~Zaldarriaga,
  JCAP {\bf 0605}, 004 (2006)
  [astro-ph/0509029].
  
\bibitem{Yadav:2007ny} 
  A.~P.~S.~Yadav, E.~Komatsu, B.~D.~Wandelt, M.~Liguori, F.~K.~Hansen and S.~Matarrese,
  Astrophys.\ J.\  {\bf 678}, 578 (2008)
  [arXiv:0711.4933 [astro-ph]].


\bibitem{Liguori:2003mb} 
  M.~Liguori, S.~Matarrese and L.~Moscardini,
  Astrophys.\ J.\  {\bf 597}, 57 (2003)
  [astro-ph/0306248].
  
\bibitem{Liguori:2007sj} 
  M.~Liguori, A.~Yadav, F.~K.~Hansen, E.~Komatsu, S.~Matarrese and B.~Wandelt,
  Phys.\ Rev.\ D {\bf 76}, 105016 (2007)
  [Erratum-ibid.\ D {\bf 77}, 029902 (2008)]
  [arXiv:0708.3786 [astro-ph]].
    
\bibitem{Elsner:2009md} 
  F.~Elsner and B.~D.~Wandelt,
  Astrophys.\ J.\ Suppl.\  {\bf 184}, 264 (2009)
  [arXiv:0909.0009 [astro-ph.CO]].

\bibitem{Oh:1998sr} 
  S.~P.~Oh, D.~N.~Spergel and G.~Hinshaw,
  Astrophys.\ J.\  {\bf 510}, 551 (1999)
  [astro-ph/9805339].

\bibitem{Smith:2007rg} 
  K.~M.~Smith, O.~Zahn and O.~Dore,
  Phys.\ Rev.\ D {\bf 76}, 043510 (2007)
  [arXiv:0705.3980 [astro-ph]].

\bibitem{Lewis:1999bs} 
  A.~Lewis, A.~Challinor and A.~Lasenby,
  Astrophys.\ J.\  {\bf 538}, 473 (2000)
  [astro-ph/9911177].

\bibitem{Jarosik:2010iu} 
  N.~Jarosik, C.~L.~Bennett, J.~Dunkley, B.~Gold, M.~R.~Greason, M.~Halpern, R.~S.~Hill and G.~Hinshaw {\it et al.},
  Astrophys.\ J.\ Suppl.\  {\bf 192}, 14 (2011)
  [arXiv:1001.4744 [astro-ph.CO]].

\bibitem{Gold:2010fm} 
  B.~Gold, N.~Odegard, J.~L.~Weiland, R.~S.~Hill, A.~Kogut, C.~L.~Bennett, G.~Hinshaw and J.~Dunkley {\it et al.},
  Astrophys.\ J.\ Suppl.\  {\bf 192}, 15 (2011)
  [arXiv:1001.4555 [astro-ph.GA]].

\bibitem{Gorski:2004by} 
  K.~M.~Gorski, E.~Hivon, A.~J.~Banday, B.~D.~Wandelt, F.~K.~Hansen, M.~Reinecke and M.~Bartelman,
  Astrophys.\ J.\  {\bf 622}, 759 (2005)
  [astro-ph/0409513].

\bibitem{Liguori:2008vf} 
  M.~Liguori and A.~Riotto,
  Phys.\ Rev.\ D {\bf 78}, 123004 (2008)
  [arXiv:0808.3255 [astro-ph]].

\bibitem{Sasaki:1995aw} 
  M.~Sasaki and E.~D.~Stewart,
  Prog.\ Theor.\ Phys.\  {\bf 95}, 71 (1996)
  [astro-ph/9507001].

\bibitem{Lyth:2004gb} 
  D.~H.~Lyth, K.~A.~Malik and M.~Sasaki,
  JCAP {\bf 0505}, 004 (2005)
  [astro-ph/0411220].

\bibitem{Turner:1985si} 
  M.~S.~Turner,
  Phys.\ Rev.\ D {\bf 33}, 889 (1986).

\bibitem{Lyth:2007jh} 
  D.~H.~Lyth,
  JCAP {\bf 0712}, 016 (2007)
  [arXiv:0707.0361 [astro-ph]].

\bibitem{Hiramatsu:2010yu} 
  T.~Hiramatsu, M.~Kawasaki, T.~Sekiguchi, M.~Yamaguchi and J.~'i.~Yokoyama,
  Phys.\ Rev.\ D {\bf 83}, 123531 (2011)
  [arXiv:1012.5502 [hep-ph]].

\bibitem{Hiramatsu:2012gg} 
  T.~Hiramatsu, M.~Kawasaki, K.~'i.~Saikawa and T.~Sekiguchi,
  Phys.\ Rev.\ D {\bf 85}, 105020 (2012)
  [arXiv:1202.5851 [hep-ph]].
      
\end{thebibliography}
\end{document}